\documentclass[twoside,twocolumn]{article}

\usepackage{blindtext} % Package to generate dummy text throughout this template 

\usepackage[sc]{mathpazo} % Use the Palatino font
\usepackage[T1]{fontenc} % Use 8-bit encoding that has 256 glyphs
\usepackage{microtype} % Slightly tweak font spacing for aesthetics

\usepackage[english]{babel} % Language hyphenation and typographical rules

\usepackage[hmarginratio=1:1,top=32mm,columnsep=20pt]{geometry} % Document margins
\usepackage[hang, small,labelfont=bf,up,textfont=it,up]{caption} % Custom captions under/above floats in tables or figures
\usepackage{booktabs} % Horizontal rules in tables

\usepackage{lettrine} % The lettrine is the first enlarged letter at the beginning of the text

\usepackage{enumitem} % Customized lists
\setlist[itemize]{noitemsep} % Make itemize lists more compact

\usepackage{abstract} % Allows abstract customization
\usepackage{titlesec} % Allows customization of titles
\renewcommand\thesection{\Roman{section}} % Roman numerals for the sections
\renewcommand\thesubsection{\roman{subsection}} % roman numerals for subsections
\titleformat{\section}[block]{\large\scshape\centering}{\thesection.}{1em}{} % Change the look of the section titles
\titleformat{\subsection}[block]{\large}{\thesubsection.}{1em}{} % Change the look of the section titles

\usepackage{fancyhdr} % Headers and footers
\usepackage{titling} % Customizing the title section

\usepackage{hyperref} % For hyperlinks in the PDF

\usepackage{scalerel, amssymb}

\usepackage{amsmath}
\usepackage{framed}
\usepackage{graphicx} %% for loading jpg figures
\usepackage{hyperref}   % to set up hyperlinks
\usepackage{multicol}
\usepackage{float}
\usepackage{nomencl}
\usepackage{wasysym}
\hypersetup{
	colorlinks=true,
	linkcolor=blue,
	citecolor=blue,
	urlcolor=blue,
}
\usepackage[square,numbers]{natbib}

\pretitle{\begin{center}\huge\bfseries} % Article title formatting
\posttitle{\end{center}} % Article title closing formatting
\title{Static force characteristic of annular gaps - \\
Experimental and simulation results} % Article title
\author{
\textsc{Maximilian M. G. Kuhr}\thanks{Corresponding author} \\[1ex]
\normalsize Chair of Fluid Systems \\
\normalsize Technische Universität Darmstadt \\
\normalsize \href{mailto:maximilian.kuhr@fst.tu-darmstadt.de}{maximilian.kuhr@fst.tu-darmstadt.de}
\and 
\textsc{Sebastian R. Lang} \\[1ex]
\normalsize KSB SE \& Co. KGaA \\
\normalsize \href{mailto:sebastian.lang@ksb.com}{sebastian.lang@ksb.com}
\and 
\textsc{Peter F. Pelz} \\[1ex]
\normalsize Chair of Fluid Systems \\
\normalsize Technische Universität Darmstadt \\
\normalsize \href{mailto:peter.pelz@fst.tu-darmstadt.de}{peter.pelz@fst.tu-darmstadt.de}
}
\date{} % Leave empty to omit a date
\renewcommand{\maketitlehookd}{%
\begin{abstract}
We discuss the static force characteristic of annular gaps resulting from an axial flow component. So far there is a severe lack of understanding of the flow inside the annulus. First, the state-of-the-art modelling approaches to describe the flow inside the annulus are recapped and discussed. The discussion focuses in particular on the modelling of inertia effects. Second, a new calculation method, the Clearance-Averaged Pressure Model (CAPM) is presented. The CAPM uses an integro-differential approach in combination with power law ansatz functions for the velocity profiles and a Hirs' model to calculate the resulting pressure field. Third, for experimental validation, a setup is presented using magnetic bearings to inherently measure the position as well as the force on the rotor induced by the flow field inside the gap. The experiments focus on the characteristic load behaviour, attitude angle and pressure difference across the annulus. Fourth, the experimental results are compared to the calculation results.
\end{abstract}
}

\begin{document}

% Print the title
\maketitle

%----------------------------------------------------------------------------------------
%	ARTICLE CONTENTS
%----------------------------------------------------------------------------------------

\section{Introduction}
In modern turbomachinery such as centrifugal pumps the reliability and performance is often limited by the dynamic behaviour of the machine, i.e. mechanical vibrations. These mechanical vibrations are highly influenced by hydrodynamic forces of the flow acting on the surface of narrow annular gaps. In general, the fluid flow within an annulus is three dimensional. The circumferential flow component driven by viscous forces is superimposed by an axial flow component caused by an axial pressure difference. Due to design and operation parameters, the flow at the gap entrance is also superimposed by a pre-swirl, which is convected into the gap by the axial flow component. So far, there is a severe lack of understanding the static and dynamic characteristics of this flow, i.e. time efficient state of the art simulation methods like the Reynolds equation of lubrication theory or the bulk-flow approach fail to reliably predict the induced forces on the rotor. This model uncertainty is crucial for the reliability and design of modern turbomachinery. 

\begin{figure}
	\centering
	\includegraphics[scale=1.0]{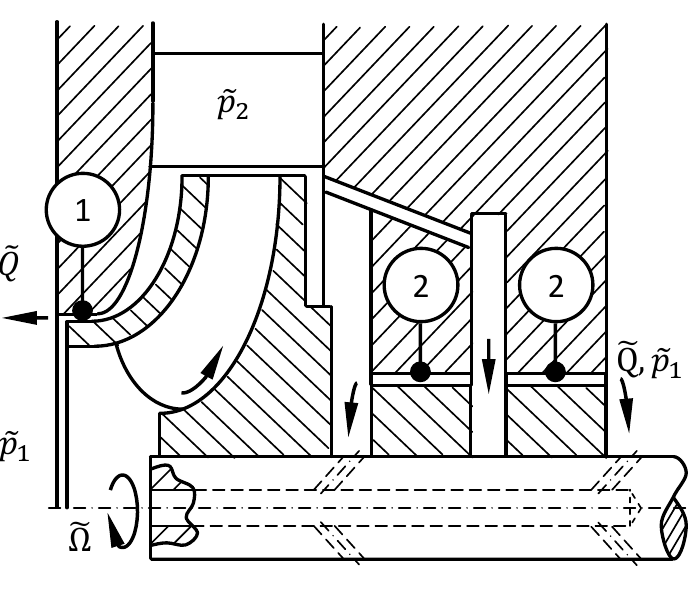}
	\caption{Schematic drawing of a modern centrifugal pump with narrow annuli: (1) annular seal; (2) media lubricated journal bearing.}
	\label{fig:introduction_modern_turbopump}
\end{figure}

Figure \ref{fig:introduction_modern_turbopump} shows a modern centrifugal pump with two narrow annular gaps exerting hydrodynamic forces on the shaft: first, a contactless annular seal or damper seal (1) and second, a media lubricated journal bearing or its oil lubricated equivalent (2).

The function of the different machine elements are "carrying a specific load $\tilde{F}$" for the journal bearing and to "seal a pressure difference" $\Delta \tilde{p} = \tilde{p}_2 - \tilde{p}_1$ for the annular seal. Nowadays, the flow within the annuli are calculated either by Reynolds' equation of lubrication theory \citep{Reynolds.1886} or the bulk-flow approach based on the work of Black \& Jenssen \citep{Black.1969}, Nelson \citep{Nelson.1985} and \citep{Childs.1993}.\\
Due to modern specifications and design regulations, the once sharp distinction between one element and the other blurs and the two elements are no longer distinguished easily. This also applies to the calculation methods used to describe either the classical journal bearing or the annular seal. This means that annular seals also have to deal with operation conditions typical for journal bearings and vice versa. \\
The state of the art calculation methods fail to predict the flow in these hybrid annuli due to the fact that either the bearing is now superimposed with an axial flow component under predominantly turbulent flow conditions or the annular seal is operated at a low pressure difference at high off-centred, i.e. eccentric, positions. To overcome this, a simulation method based on the work of Lang \citep{Lang.2018} is presented in this paper combining the possibility to calculate the static force characteristic for hybrid annular gaps with an axial flow component.\\ 
The presented paper is structured into three parts: (i) first, the state of the art simulation methods for narrow annuli are recapitulated. (ii) Second, a new an general method is presented using the momentum and continuity equation in combination with ansatz functions and a Hirs' model \citep{Hirs.1973} to calculate the resulting pressure field. (iii) Third, the experimental setup to determine the characteristic behaviour is presented and the experimental investigations and calculation results are compared and discussed.

\subsection{The effect of fluid inertia in annular gap flow}
\begin{figure*}
	\centering
	\includegraphics[scale=1.0]{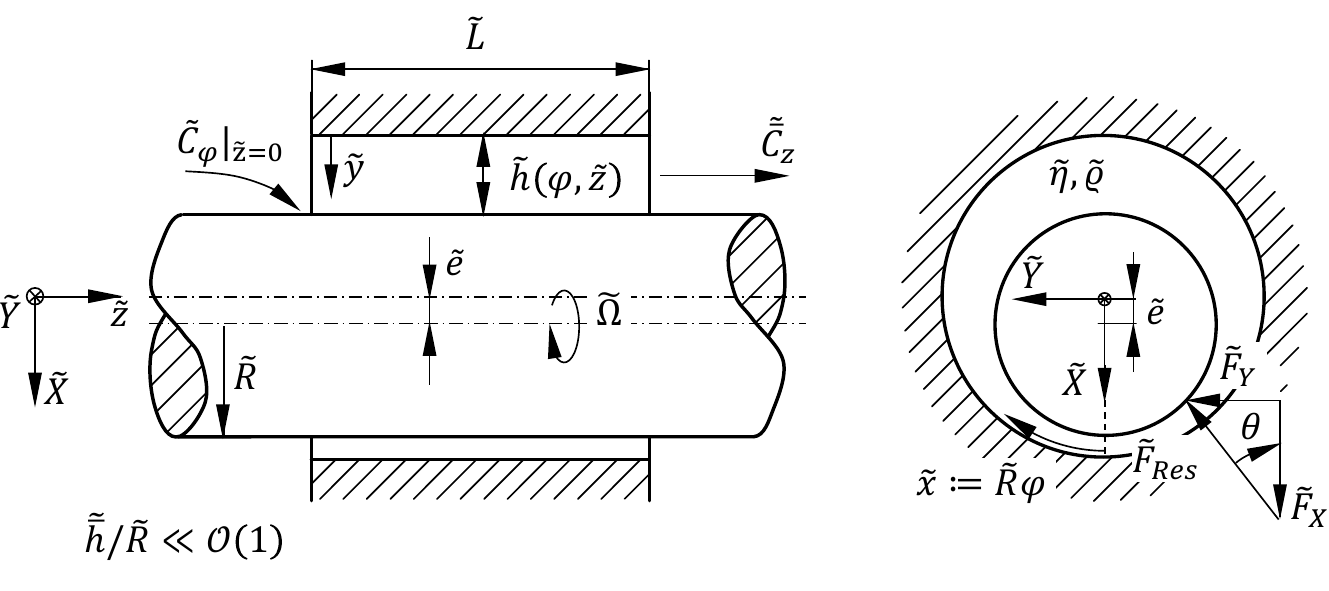}
	\caption{Generic annular gap with an axial flow component.}
	\label{fig:introduction_generic_annulus}
\end{figure*}

Figure \ref{fig:introduction_generic_annulus} shows an eccentrically operated generic narrow annular gap. The geometry is given by the shaft radius $\tilde{R}$, the annulus length $\tilde{L}$ and the resulting gap function $\tilde{h}(\tilde{x},z)$. The mean gap height $\tilde{\bar{h}}$ divided by the shaft radius $\tilde{R}$ is much smaller than one, i.e $\tilde{\bar{h}} / \tilde{R} \ll 1$. The operating conditions are given by the eccentricity $\tilde{e}$ of the shaft relative to the stationary wall, the angular frequency $\tilde{\Omega}$ of the shaft, the mean axial velocity $\tilde{\bar{C}}_z$ and the pre-swirl velocity at the entrance of the annulus $\tilde{C}_{\varphi} \vert_{\tilde{z} = 0}$. The fluid properties are given by the density $\tilde{\varrho}$ and the dynamic viscosity $\tilde{\mu}$. The induced pressure field induces the resulting force component $\tilde{F}_{Res}$ under an attitude angle $\theta$. Here, the tilde $\tilde{\square}$ declares variables with dimensions, whereas dimensionless variables are written without it.

On dimensional ground the induced dimensionless pressure $p := 2 \tilde{p} / ( \tilde{\varrho} \tilde{\Omega}^2 \tilde{R}^2 )$ is only a function of six dimensionless measures: (i) the relative gap clearance $\psi := \tilde{\bar{h}} / \tilde{R}$, (ii) the dimensionless annulus length $L := \tilde{L} / \tilde{R}$, (iii) the relative eccentricity $\varepsilon := \tilde{e} / \tilde{\bar{h}}$, (iv) the Reynolds number in circumferential direction $Re_\varphi := \tilde{\Omega} \tilde{R} \tilde{\bar{h}} / \tilde{\nu}$, (v) the flow number $\phi := \tilde{\bar{C}}_z / ( \tilde{\Omega} \tilde{R} )$ and (vi) the pre-swirl $C_\varphi \vert_{z=0} := \tilde{C}_\varphi \vert_{z=0} / (\tilde{\Omega} \tilde{R})$:
\begin{equation}
	p = p \left( \psi, L, \varepsilon, Re_\varphi, \phi, C_\varphi \vert_{z=0} \right).
\end{equation}

So far we made the distinction between seals and bearing with the difference in function, i.e. sealing a pressure difference vs. carrying a specific load. From a fluid mechanics or physical perspective, whether or not the inertia of the fluid is relevant is important to the method we present in this paper.  

To distinguish whether or not inertia has to be taken into account an order of magnitude analysis leads to a characteristic parameter $\psi Re_\varphi$, cf. \citep{Kahlert.1948,Spurk.2008,Lang.2016,Lang.2018}. This modified Reynolds number indicates the influence of inertia in relation to viscous forces and divides the flow into three different categories: (i) $\psi Re_\varphi \ll 1$ fluid inertial is negligible compared to viscous forces; (ii) $\psi Re_\varphi \sim 1$ both viscous and convective terms need to be considered within the calculation and (iii) $\psi Re_\varphi \gg 1$ inertia is dominant compared to viscous forces. 

Initially focusing on the annular seal, the relative clearance and the Reynolds number are of order $\psi \sim 10^{-2}$ and $Re_\varphi \sim 10^3...10^5$, resulting in the characteristic parameter $\psi Re_\varphi \sim 10^1...10^3 \gg 1$. This yields that the flow within an annular seal is dominated by inertia effects. To further distinguish between laminar and turbulent flow Spurk \& Aksel \citep{Spurk.2008} formulate a critical Reynolds number $Re_\mathrm{cirt} := 41.3 \sqrt{1 / \psi} \sim 10^{2}$. If this limit is exceeded the former pure laminar flow is disturbed by Taylor vortices. A further increase in Reynolds number then leads to the formation of turbulent flow. In modern annular seals, the Reynolds numbers are always much larger than the critical Reynolds number, indicating completely turbulent flow within the annulus. 

Now focusing on classical oil lubricated journal bearings, the relative clearance is of order $\psi \sim 10^{-3}$. Due to high fluid viscosity the corresponding Reynolds number are of order $Re_\varphi \sim 10^0...10^1$, resulting in the characteristic parameter $\psi Re_\varphi \sim 10^{-2}...10^{-1} \ll 1$. This indicates that the flow within a classic journal bearing is usually dominated by viscous effects. The Reynolds number is often smaller than the critical Reynolds number $Re_\mathrm{cirt} \sim 10^3$ indicating laminar flow within the bearing. 

In general, however, such an asymptotic distinction between seal or bearing is inappropriate in modern turbomachinery. Due to low viscosity lubricants such as water or air as well as larger relative clearances and the presence of significant axial pressure differences the characteristic parameter for most annuli is of the order $\psi Re_\varphi \sim 1$, which necessitates the inclusion of convective terms in the model. In addition, the Reynolds number can either be below, equal or above the critical Reynolds number, i.e. $Re_\varphi \leq Re_\mathrm{cirt}$ or $Re_\mathrm{cirt}  < Re_\varphi$. Thus, both laminar and turbulent flow depending on the operation point as well as the shape and size, i.e. the geometry, of the annulus. From a physical point of view, these reasons indicate a continuous transition between the bearing and the annulus seal.  

\subsection{The bulk-flow model}
The existing literature on the static and dynamic force characteristics of bearings and seals is based either on the assumption $\psi Re_\varphi \ll 1$, neglecting the left-hand side of the equation of motion, i.e. any inertia terms, or $\psi Re_\varphi\gg 1$ neglecting any viscous friction forces. 

Despite complex transient 3D CFD Simulations, annular seals are in general calculated using an integro-differential approach, cf. \citep{Launder.1978,Launder.1978b}. The integro-differential approach is obtained by integrating the Navier-Stokes equation over the annulus height $\tilde{h}$ assuming an incompressible flow without volume forces. Using Cartesian index notation, this yields 
\begin{equation}
\begin{split}
	&\int_{0}^{\tilde{h}}\frac{\partial \tilde{c}_j}{\partial \tilde{x}_j}\,\mathrm{d}\tilde{y} = 0,\\
	&\int_{0}^{\tilde{h}}\tilde{\varrho}\tilde{c}_j\frac{\partial \tilde{c}_i}{\partial \tilde{x}_j}\,\mathrm{d}\tilde{y} = -\tilde{h}\frac{\partial \tilde{p}}{\partial \tilde{x}_i} + \tilde{\tau}_{yi}|^{\tilde{h}}_{0}, \quad i,j=1,2.
\end{split}
\end{equation}

Using the dimensionless variables and averaging the velocity profiles over the gap height $\bar{C}_i := 1 / h \int_{0}^{h} c_i \, \mathrm{d}y$ yields the so called bulk-flow model based on the work of Black \& Jenssen \citep{Black.1969}, Nelson \citep{Nelson.1985} and \citep{Childs.1993}
\begin{equation}
    \begin{split}
        &\frac{\partial (h\bar{C}_\varphi)}{\partial \varphi} + \frac{\phi}{L} \frac{\partial (h\bar{C}_z)}{\partial z},\\
        &h \bar{C}_\varphi \frac{\partial \bar{C}_\varphi}{\partial \varphi} + \frac{\phi}{L} h \bar{C}_z \frac{\partial \bar{C}_\varphi}{\partial z} = -\frac{h}{2}\frac{\partial p}{\partial \varphi} + \tau_{y\varphi},\\
        &\phi h \bar{C}_\varphi \frac{\partial \bar{C}_z}{\partial \varphi} + \frac{\phi^2}{L} h \bar{C}_z \frac{\partial \bar{C}_z}{\partial z} = -\frac{h}{2L}\frac{\partial p}{\partial z} + \tau_{yz}.
    \end{split}
\end{equation}

The bulk-flow model has been used by various authors, cf. Childs \citep{Childs.1983,Childs.1983b}, San Andr\'{e}s \citep{SanAndres.1990,SanAndres.1991,SanAndres.1995}, Nelson \& Nguyen \citep{Nelson.1988,Nelson.1988b}, and has become the state of the art calculation method for annular gaps and hydrostatic bearings. However, the bulk-flow model is by no means uncritical: Hirs \citep{Hirs.1973} and Szeri \citep{Szeri.1998} criticise the use of gap height averaged velocity profiles. Strictly speaking, the assumption is only valid for block-shaped velocity profiles, i.e. $Re_\varphi \to \infty$. To overcome this drawback, Launder \& Leschziner \citep{Launder.1978,Launder.1978b} as well as Simon \& Fr\^{e}ne \citep{Simon.1992} use parabolic ansatz functions $\int_{0}^{h}c^2_i \, \mathrm{d}y := a C_i^2 + b C_i + d$ to describe the integrals of the velocity profiles within the annulus. Here, the disadvantage lies in the complex calibration of the describing coefficients $a$, $b$ and $d$. Additional literature as well as a detailed overview of published articles using either CFD simulations or the integro-differential approach is found in Lang \citep{Lang.2018}.
 
\subsection{Lubrication theory}
Journal bearings with either laminar or turbulent flow are usually treated by Reynolds' differential equation of hydrodynamic lubrication theory \citep{Reynolds.1886}. Subjecting the Navier-Stokes equation to an order of magnitude analysis, it is easy to show that the fluid inertia represented by the convective terms on the left hand side of the momentum equation is negligible small if $\psi Re_\varphi \ll 1 $. This yields
\begin{equation}
    \frac{\partial}{\partial \varphi} \left( \frac{h^3}{k_\varphi} \frac{\partial p}{\partial \varphi} \right) + \frac{1}{L^2} \left( \frac{1}{k_z} \frac{\partial p}{\partial z} \right) = \frac{1}{\psi Re_\varphi} \frac{\partial h}{\partial \varphi} + I(c_\varphi, c_z).
\end{equation}

For turbulent flow within the annulus ($\psi Re_\varphi \sim 1 $) convective terms are still neglected and the fluid inertia, e.g. turbulence effects, are modelled by adding empirical correction coefficients $k_\varphi,\,k_z$ as well as a source term $I(c_\varphi, c_z)$ depending on the fluid velocity in circumferential and axial direction. The use of the coefficients $k_\varphi$ and $k_z$ can be interpreted as a local correction of the fluid viscosity and is inspired by the general turbulence modelling of viscous stresses, cf. \citep{Pope.2015}. In general the correction coefficients depend on the Reynolds number $Re_\varphi$ of the corresponding operating point, cf. \citep{Constantinescu.1959, Constantinescu.1982, Hirs.1973, Simon.1989}. Exemplary, the correction coefficients of Constantinescu \& Galetuse \citep{Constantinescu.1982} are given by
\begin{equation}
    \begin{split}
        &k_\varphi = 12 + 0.0136 Re_\varphi^{0.9},\\
        &k_z = 12 + 0.0043 Re_\varphi^{0.96},\\
        &k_\varphi, k_z = 12 \; \mathrm{for} \; Re_\varphi<1000. 
    \end{split}
\end{equation}

The additional source term results out of the convective terms if they are assumed to be relevant instead of negligible \citep{Constantinescu.1970,Constantinescu.1974,Constantinescu.1982}. In contrast to the standard Reynolds' differential equation the added source term leads to numerical problems regarding solution convergence, cf. \citep{Constantinescu.1982}. 

This asymptotic view paired with the mentioned drawbacks and the fact that publications dealing with generic annular gaps with an axial flow component containing both calculation results and experimental investigations of one single author are limited leads to a need for new, experimentally validated and generalised calculation methods, cf. \citep{Lang.2018}. 
\section{The Clearance-Averaged Pressure Model - CAPM}\label{sec:CAPM}
To overcome those drawbacks a new calculation method is presented. Based on the work of Lang \citep{Lang.2016,Lang.2018} and Robrecht et al. \cite{Robrecht.2019}, the Clearance-Averaged Pressure Model (CAPM) uses an integro-differential approach to describe the flow within the annulus. In contrast to the bulk-flow model, the integrals in the governing equations are treated by using power law ansatz function for the velocity profiles. In the following, the basic equations and the boundary conditions for calculating the pressure field inside the annulus are derived. Furthermore, an analysis of the describing equations leads to a further reduction of dimensionless variables. 
\begin{figure}
	\centering
	\includegraphics[scale=1.0]{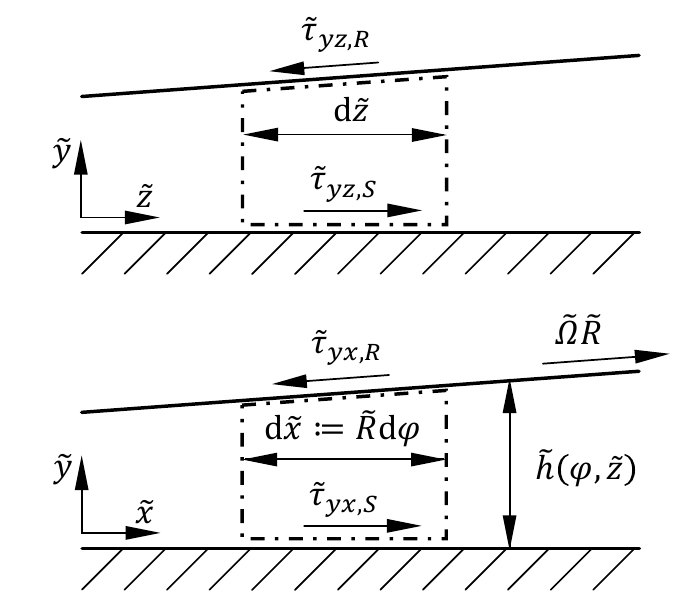}
	\caption{Control volume within the annulus. Upper figure shows the control volume in axial direction, whereas the lower figure shows the control volume in circumferential direction.}
	\label{fig:numerical_setup_control_volume}
\end{figure}

Figure \ref{fig:numerical_setup_control_volume} shows the control volume of the generic annuls. Due to dimensional reasons, the radial velocity component as well as the pressure gradient over the gap height are negligible. The stationary dimensionless continuity and momentum equation in circumferential and axial direction yields the system of non-linear partial differential equations 
\begin{equation}
\begin{split}
&\frac{\partial}{\partial \varphi}h\int_{0}^{1}c_\varphi\,\mathrm{d}y + \frac{\phi}{L} \frac{\partial}{\partial z}h\int_{0}^{1}c_z\,\mathrm{d}y = 0, \\
&\frac{\partial}{\partial \varphi}h\int_{0}^{1}c^2_\varphi\,\mathrm{d}y + \frac{\phi}{L} \frac{\partial}{\partial z}h\int_{0}^{1}c_\varphi c_z\,\mathrm{d}y = \\
&\quad\quad\quad\quad\quad\quad\quad\quad\quad\quad -\frac{h}{2}\frac{\partial p}{\partial \varphi} + \frac{1}{2\psi}\tau_{y\varphi}\big|^1_0,\\%\frac{\tau_{\mathrm{stat},\varphi} - \tau_{\mathrm{rot},\varphi}}{2\psi},\\
&\phi \frac{\partial}{\partial \varphi} h\int_{0}^{1}c_\varphi c_z\,\mathrm{d}y + \frac{\phi^2}{L}\frac{\partial}{\partial z}h\int_{0}^{1}c^2_z\,\mathrm{d}y = \\
&\quad\quad\quad\quad\quad\quad\quad\quad\quad\quad  -\frac{h}{2L}\frac{\partial p}{\partial z} + \frac{1}{2\psi}\tau_{yz}\big|^1_0 %+ \frac{\tau_{\mathrm{stat},z} - \tau_{\mathrm{rot},z}}{2\psi}.
\end{split}
\end{equation}

The integrals in the equations are solved analytically by using power law ansatz functions to describe the velocity profiles in circumferential and axial direction, cf. \citep{Lang.2018}. Due to the modular integration of the velocity profiles, the ansatz functions can be adapted to gap flows at arbitrary Reynolds numbers. The requirement $Re_\varphi \to \infty$ for block-shaped velocity profiles in the bulk-flow approach is thus eliminated. In particular, the power law ansatz functions can be used to treat both laminar and turbulent gap flows in a unified model framework. In the literature, the velocity profiles used for the ansatz functions have so far always been assumed to be "unaffected by inertial effects", i.e. fully developed velocity profiles are always used to evaluate the velocity integrals, cf. \citep{Launder.1978,Launder.1978b,SanAndres.1991,Simon.1992}. The development of the velocity profiles through the annulus starting from block-shaped velocity profiles at the gap inlet to parabolic velocity profiles downstream has not been considered so far. Initially, however, this step will not be pursued further in this paper. However, due to the modular integration of the velocity profiles, ansatz functions are suitable for the consideration of such inlet processes. Instead, the possibility of modelling the velocity profiles using power law ansatz functions assuming fully developed velocity profiles will be investigated. This is reasonable if the typical annulus length $L$ is much larger than the hydrodynamic entrance length $L_{hyd} := \tilde{L}_{hyd} / \tilde{R}$, i.e. $L \gg L_{hyd}$. Stampa \citep{Stampa.1971} and Herwig \citep{Herwig.2015} show that the hydraulic entrance length is a power law function of the Reynolds number $Re_\varphi$, the flow number $\phi$, the relative gap clearance $\psi$ and an empirical constant $k_{hyd}$ 
\begin{equation}
    L_{hyd} = k_{hyd} \, \psi \left(Re_\varphi \sqrt{ 1 + \phi ^2} \right)^{n_{hyd}}.
\end{equation}
Both authors report empirical constants in the range of $0.66 < k_{hyd} < 8.8$ and power law exponents $1/6 < n_{hyd} < 1/4$. Considering typical Reynolds numbers being in the order of magnitude of $Re_\varphi  \sim 10^3 ... 10^5$ as well as flow numbers being in the order of magnitude $\phi \sim 1$ and typical relative clearances being in the order of magnitude $\psi \sim 10^{-3}$, the hydrodynamic entrance length is in the order of magnitude $L_{hyd} \sim 10^{-3} ... 10^{-1}$. This is good agreement with the numerical results obtained by Lang \citep{Lang.2018}.
% Within this paper we focus on generic annular gaps with an axial flow component and turbulent flow conditions. Therefore, the power law ansatz functions reads 
Within this paper we focus on generic annular gaps with lengths being in the order of magnitude $L \sim 1$, i.e. it is reasonable to assume fully developed velocity profiles. Therefore, the power law ansatz functions reads 
\begin{equation}
    \begin{split}
        & c_\varphi := \begin{cases}
                \left( C_\varphi - 1 \right) \left[ 2 \left( y_R \right) \right]^{1/n_\varphi} + 1 & \; \, \text{for } y_R < 1/2 \\
                C_\varphi \left( 2 \, y \right)^{1/n_\varphi} & \; \, \text{for } y \le 1/2
            \end{cases},\\
        & c_z := \begin{cases}
                C_z \left[ 2 \left( y_{\mathrm{ROT}} \right) \right]^{1/n_z} & \qquad\qquad \text{for } y_R < 1/2 \\
                C_z \left( 2 \, y \right)^{1/n_z} &  \qquad\qquad \text{for } y \le 1/2
            \end{cases}.
    \end{split}   
\end{equation}
For the sake of simplicity, the coordinate $y_R := 1 - y$ is introduced starting at the rotor surface. $C_\varphi$ and $C_z$ are the centreline velocities at half gap height and $n_\varphi = 5, n_z = 6.5$ the exponents of the power law ansatz functions. The exponents were obtained by an extensive numerical study using a three-dimensional CFD model with approximately 10.9 million cells and an RSM turbulence model. The model was solved using the commercial software ANSYS Fluent, resulting in a good comparison to the exponents reported by Sigloch \citep{Sigloch.2017} and Reichardt \citep{Reichardt.1959}. Therefore, the integrals for the continuity and the momentum equations yield
\begin{equation}
\begin{split}
& \int_{0}^{1}c_\varphi\,\mathrm{d}y = \frac{n_\varphi}{n_\varphi + 1} C_\varphi + \frac{1}{2\left(n_\varphi + 1 \right)},\\
& \int_{0}^{1}c_z\,\mathrm{d}y = \frac{n_z}{n_z + 1} C_z,\\
& \int_{0}^{1}c^2_\varphi\,\mathrm{d}y = \frac{n_\varphi}{\left(n_\varphi+2\right)\left(n_\varphi+1\right)}C_\varphi + \\ 
& \quad\quad\quad\quad\quad\quad + \frac{n_\varphi}{n_\varphi +2}C^2_\varphi + \frac{1}{\left(n_\varphi+2\right)\left(n_\varphi+1\right)},\\
& \int_{0}^{1}c_\varphi c_z\,\mathrm{d}y = \frac{n_\varphi n_z}{n_\varphi n_z + n_\varphi + n_z}C_z C_\varphi + \\
& \quad\quad\quad\quad\quad\quad + \frac{n_z^2}{2\left(n_\varphi n_z + n_\varphi + n_z\right)\left(n_z + 1\right)}C_z,\\
& \int_{0}^{1}c^2_z\,\mathrm{d}y = \frac{n_z}{n_z + 2}C^2_z.
\end{split}
\end{equation}

The wall shear stresses $\tau_{yi}\big|^1_0$ are modelled according to the bulk-flow theory for turbulent film flows. Separating the wall shear stresses $\tau_{yi}\big|^1_0$ into their corresponding directional part, cf. figure \ref{fig:numerical_setup_control_volume}, yields
\begin{equation}
\tau_{yi}\big|^1_0 = \tau_{yi,S} - \tau_{yi,R}.
\end{equation}

The directional wall shear stresses $\tau_{yi,S}$ and $\tau_{yi,R}$ read,
\begin{equation}
\begin{split}
& \tau_{yi,S} = f_S C_S C_{i,S},\\
& \tau_{yi,R} = f_S C_R C_{i,R}.
\end{split}
\end{equation}

with the Fanning friction factor $f_i, i=R,S$, the dimensionless effective relative velocity between the wall (rotor $R$, stator $S$) and the fluid $C_i:=\sqrt{C^2_{\varphi,i} + \phi^2 C^2_{z,i}}, i=R,S$. Here, the components $C_{\varphi,i}$ and $C_{z,i}$ are boundary layer averaged velocities between the wall and the corresponding boundary layer thickness $\delta$ assuming fully developed boundary layers throughout the annulus. The boundary layer averaged velocities yield
\begin{equation}
\begin{split}
& C_{\varphi, S} := \frac{1}{\delta}\int_{0}^{\delta}c_\varphi\,\mathrm{d}y = \frac{n_\varphi}{n_\varphi + 1} C_\varphi,\\
& C_{\varphi, R} := \frac{1}{\delta}\int_{0}^{\delta}\left(c_\varphi-1\right)\,\mathrm{d}y = \frac{n_\varphi}{n_\varphi + 1} \left(C_\varphi-1\right),\\
& C_{z, S} := \frac{1}{\delta}\int_{0}^{\delta}c_z\,\mathrm{d}y = \frac{n_z}{n_z + 1} C_z,\\
& C_{z, R} := \frac{1}{\delta}\int_{0}^{\delta}c_z\,\mathrm{d}y = \frac{n_z}{n_z + 1} C_z.
\end{split}
\end{equation}
 
The fanning friction factor is modelled using Hirs' wall friction model
\begin{equation}
	f_i:= m_f \left(\frac{h}{2}C_i Re_\varphi\right)^{-n_f}.
\end{equation}

Here, the friction factor is modelled using the empirical constants $n_f$ and $m_f$ as well as the Reynolds number $Re_\varphi$. The empirical constants describe arbitrary lines within the double logarithmic Moody diagram. By careful selection of the coefficients, it is possible to model both laminar and turbulent friction. In addition, it is possible to model hydraulically smooth and hydraulically rough friction behaviour. Furthermore, a representation of the transition area between laminar and turbulent flow is supported. Modelling the wall shear stresses by using the fanning friction factor in combination with the Hirs' model the wall shear stresses are of the order
\begin{equation}
	\tau_{yi} \sim m_f Re_\varphi^{-n_f}.
\end{equation}  

It follows that the Reynolds number and the relative clearance occur only together in the momentum equations. Lang \citep{Lang.2018} uses this to reduce the number of depending dimensionless variables by introducing a modified Reynolds number $Re^*_\varphi := \psi Re^{n_f}_\varphi$ inspired by the characteristic parameter $\psi Re_\varphi$. For laminar flow, i.e. $n_f = 1$, the modified Reynolds number reduces to the characteristic parameter. Therefore, the induced pressure is a function of five dimensionless measures
\begin{equation}
p = p\left(L, \varepsilon, Re^*_\varphi, \phi, C_\varphi\vert_{z=0}\right).
\end{equation}

In order to solve the system of equations the boundary conditions on the gap entrance and exit have to be specified. Here, two types of boundary conditions are imposed on the annular gap. A pressure boundary condition at the inlet and the exit of the annulus. Due to the inertia of the fluid a separation bubble forms at sharp edge of the gap inlet when the flow enters the annulus. The bubble is confined to a small area forming a convergent divergent nozzle. Here, the fluid is initially accelerated, resulting in a circumferentially distributed lowered static pressure. In the divergent part of the nozzle the flow expands, resulting in a circumferential distributed static pressure as well as a Carnot loss. This phenomenon was intentionally described by Lomakin \citep{Lomakin.1958} and is widely documented within the literature, cf. \citep{Childs.1993, Brennen.1994}. The pressure loss is modelled by applying Bernoulli's equation. The pressure boundary conditions at the gap inlet read,
\begin{equation}
	\Delta p = p\big|_{z=0} + \left(1+\zeta\right)\left(\phi^2 C^2_z + C^2_\varphi\right),\\
\end{equation}

Here, $\Delta p$ is the pressure difference across the annulus and $\zeta$ describes the inlet pressure loss coefficient. In general, $\zeta$ is a function of the inlet geometry as well as the relative gap clearance and operating parameters including the Reynolds number $Re_\varphi$, the flow number $\phi$ and the relative shaft displacement, i.e. $\varepsilon$
\begin{equation}
	\zeta = \zeta\left(\mathrm{GEOM}, \psi, Re_\varphi, \phi, \varepsilon\right).
\end{equation}

Due to this complex functional relation which cannot be investigated by simple means, a constant loss coefficient is assumed throughout the paper. However, although this assumption is a severe simplification especially at high eccentricities it is often used in the literature, cf. \cite{Childs.1993, Brennen.1994, SanAndres.1991}. A Dirichlet boundary condition for the pressure is applied at the exit of the gap. For simplicity, the pressure at the gap exit reads
\begin{equation}
	p\big|_{z=0} = 0.
\end{equation}

In addition, a constant circumferential velocity boundary condition representing the pre-swirl at the annulus entrance is applied
\begin{equation}
	C_\varphi\big|_{z=0} \in \mathbb{R}.
\end{equation}  

The non-linear partial differential equation system can not be solved analytically. To solve the equation system numerically, a finite difference scheme is used. A two-di\-men\-sional uniform structured grid with a resolution of 51 x 51 cells is used. The velocities are determined at the grid nodes and the pressure is determined at the cell centre. The non-linearity in the integrals is handled by applying the Picard-iteration. The spatial derivatives are approximated by hybrid schemes. The solution of the partial differential equation system, i.e. the induced pressure filed inside the annulus $p(x,z)$ and the centreline velocities $C_\varphi, C_z$, are obtained by using a SIMPLE-C (Semi-Implicit Method for Pressure Linked Equations - Consistent) algorithm based on the work of van Doormaal \& Raithby \citep{vanDoormaal.1984}.

The induced static forces on the rotor are given by
\begin{equation}
\begin{split}
& F_X = -\int_{0}^{1}\int_{0}^{2\pi} p\, \cos\varphi \,\mathrm{d}\varphi\,\mathrm{d}z,\\
& F_Y = -\int_{0}^{1}\int_{0}^{2\pi} p\, \sin\varphi \,\mathrm{d}\varphi\,\mathrm{d}z,\\
& F_{\mathrm{Res}} = \sqrt{F^2_X + F^2_Y}, \quad \theta = \tan^{-1}\left(-\frac{F_Y}{F_X}\right).
\end{split}
\end{equation} 
 
Here, $F_X$ and $F_Y$ are the induced forces in $X$ and $Y$ direction, whereas $F_{\mathrm{Res}}$ is the resulting force and $\theta$ the attitude angle.

\section{pre-Validation}
\begin{figure*}[!ht]
	\centering
	\includegraphics[scale=1.0]{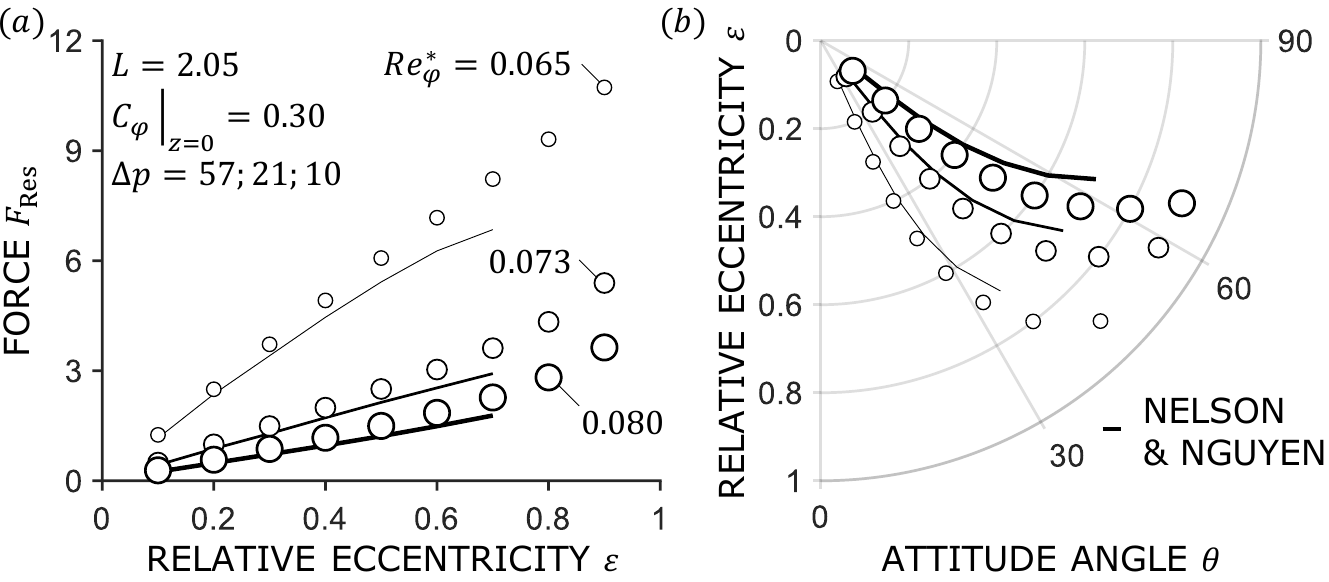}
	\caption{The (a) resulting force acting on the rotor and (b) the attitude angle determined by the CAPM compared to the results by Nelson \& Nguyen~\citep{Nelson.1988b}.}
	\label{fig:validation_Nelson1987}
\end{figure*}

\begin{figure}[!ht]
	\centering
	\includegraphics[scale=1.0]{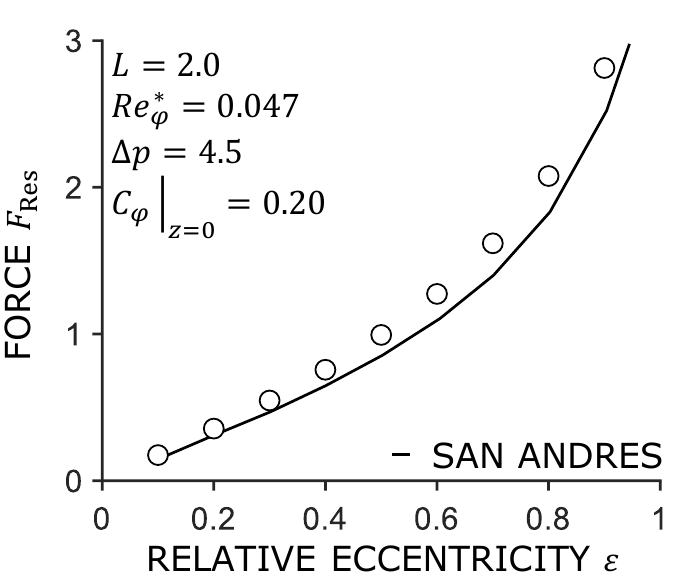}
	\caption{The resulting force acting on the rotor determined by the CAPM compared to the results by San Andr\'{e}s~\citep{SanAndres.1991b}.}
	\label{fig:validation_SanAndres1991}
\end{figure}
Prior to validating the Clearance-Averaged pressure Model using the experimental set up presented later on, the static characteristic, i.e. the resulting force $F_{\mathrm{Res}}$ on the rotor and the attitude angle $\theta$ are compared to the numerical results by Nelson \& Nguyen~\citep{Nelson.1988b} and San And\'{e}s~\citep{SanAndres.1991b}. Here, the resulting force on the rotor as well as the attitude angle is investigated. Nelson \& Nguyen use a fast Fourier transform to solve the bulk-flow equations and calculate the static characteristics of an annular seal with length $L = 2.05$, a pre-swirl $C_\varphi|_{z=0} = 0.3$ and three dimensionless pressure differences $\Delta p := 2\Delta \tilde{p} / \left( \tilde{\varrho} \tilde{\Omega}^2 \tilde{R}^2 \right) = 57;\,21\,10$ corresponding to the three modified Reynolds numbers $Re_\varphi^* = 0.065;\,0.073;\,0.080$. In contrast, San And\'{e}s uses a form of the formerly presented bulk-flow model to calculate the resulting force on the rotor of an annulus with length $L = 2$, a modified Reynolds number, a pressure difference $\Delta p = 4.5$ and a pre-swirl $C_\varphi|_{z=0}=0.3$.\\
Figure~\ref{fig:validation_Nelson1987} shows the comparison of the resulting force acting on the rotor and the attitude angle. The lines are the results by Nelson \& Nguyen~\citep{Nelson.1988b} and the markers represent the calculation results by the CAPM. Furthermore, the thickness of the lines as well as the size of the Marker correlates to the modified Reynolds number. The thicker the line thickness and the larger the  marker,  the  higher the modified Reynolds number. The calculations of the resulting force acting on the rotor $F_{\mathrm{Res}}$ as well as the attitude angle $\theta$ show a good agreement with the results obtained by Nelson \& Nguyen. The predicted forces acting on the rotor by the Clearance-Averaged Pressure Model are slightly higher than the resulting forces by Nelson \& Nguyen. Regarding the attitude angle, the trend is reserved. Here, the CAPM results are slightly below the calculations by Nelson \& Nguyen.\\
Figure~\ref{fig:validation_SanAndres1991} shows the comparison of the resulting force acting on the rotor determined by the CAPM and the data published by San And\'{e}s~\citep{SanAndres.1991b}. Due to the lack of data, a comparison of the attitude angles similar to the results by Nelson \& Nguyen is not possible. Again, the forces acting on the rotor predicted by the Clearance-Averaged Pressure Model are in good comparison to the results obtained by the bulk-flow model by San And\'{e}s. Similar to the comparison with the predictions by Nelson \& Nguyen, the CAPM results are slightly above the ones of the bulk-flow model. 

\section{The experimental setup}
\begin{figure*}
	\centering
	\includegraphics[scale=1.0]{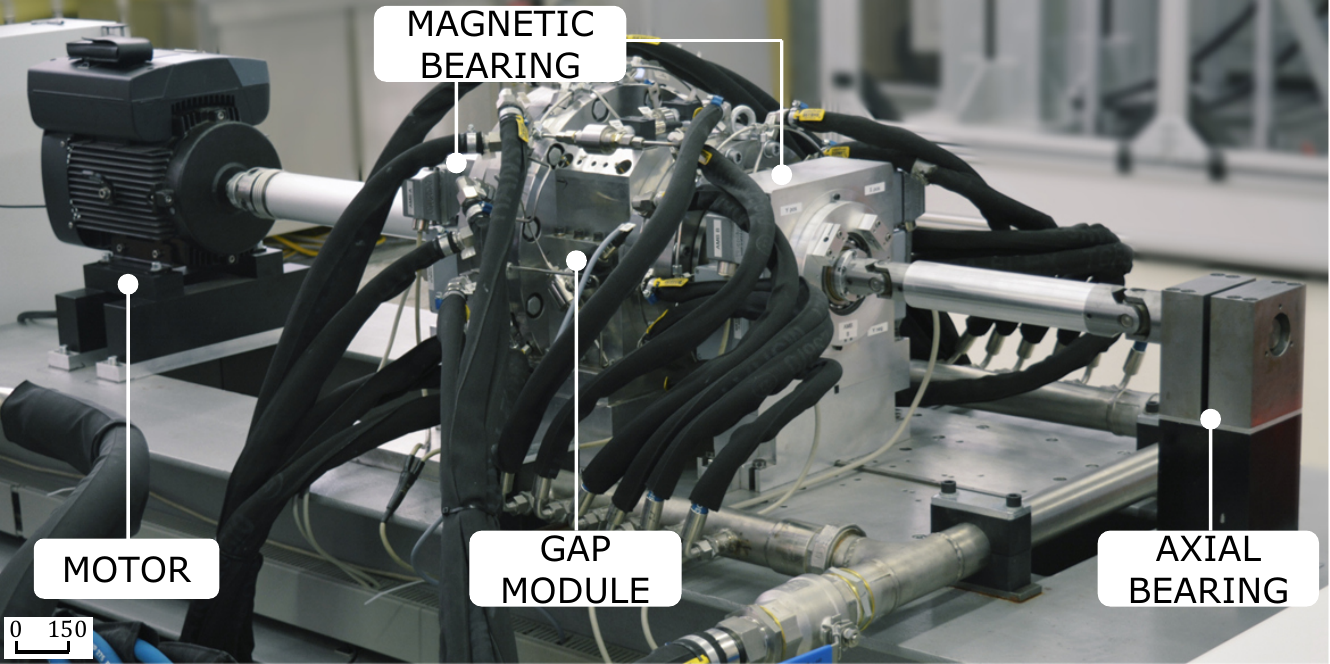}
	\caption{Photograph of the annular gap test rig at the Chair of Fluid Systems at the Technische Universität Darmstadt.}
	\label{fig:exp_test_rig}
\end{figure*}

For further validation purpose and experimental investigation of the induced forces on the rotor, a world wide unique test rig is designed. Additional information on the test rig can be found in the work of Kuhr et al. \citep{Kuhr.2019,Kuhr.2021}. Figure \ref{fig:exp_test_rig} shows the annular gap flow test bench at the laboratory of the Chair of Fluid Systems at the Technische Universität Darmstadt.

The test bench essentially consists of two magnetic bearings supporting the rotor. They also serve as an inherent displacement and force measurement system. Compared to existing test rigs where the shaft is supported by ball or journal bearings, cf. \citep{Childs.1994,Jolly.2018,Moreland.2018}, magnetic bearings have the advantage of being completely contactless and thus frictionless. In addition, the ability to displace and excite the shaft at user-defined frequencies makes them ideal for determining the static and dynamic characteristics of annular gap flows.
\begin{figure*}
	\centering
	\includegraphics[scale=1.0]{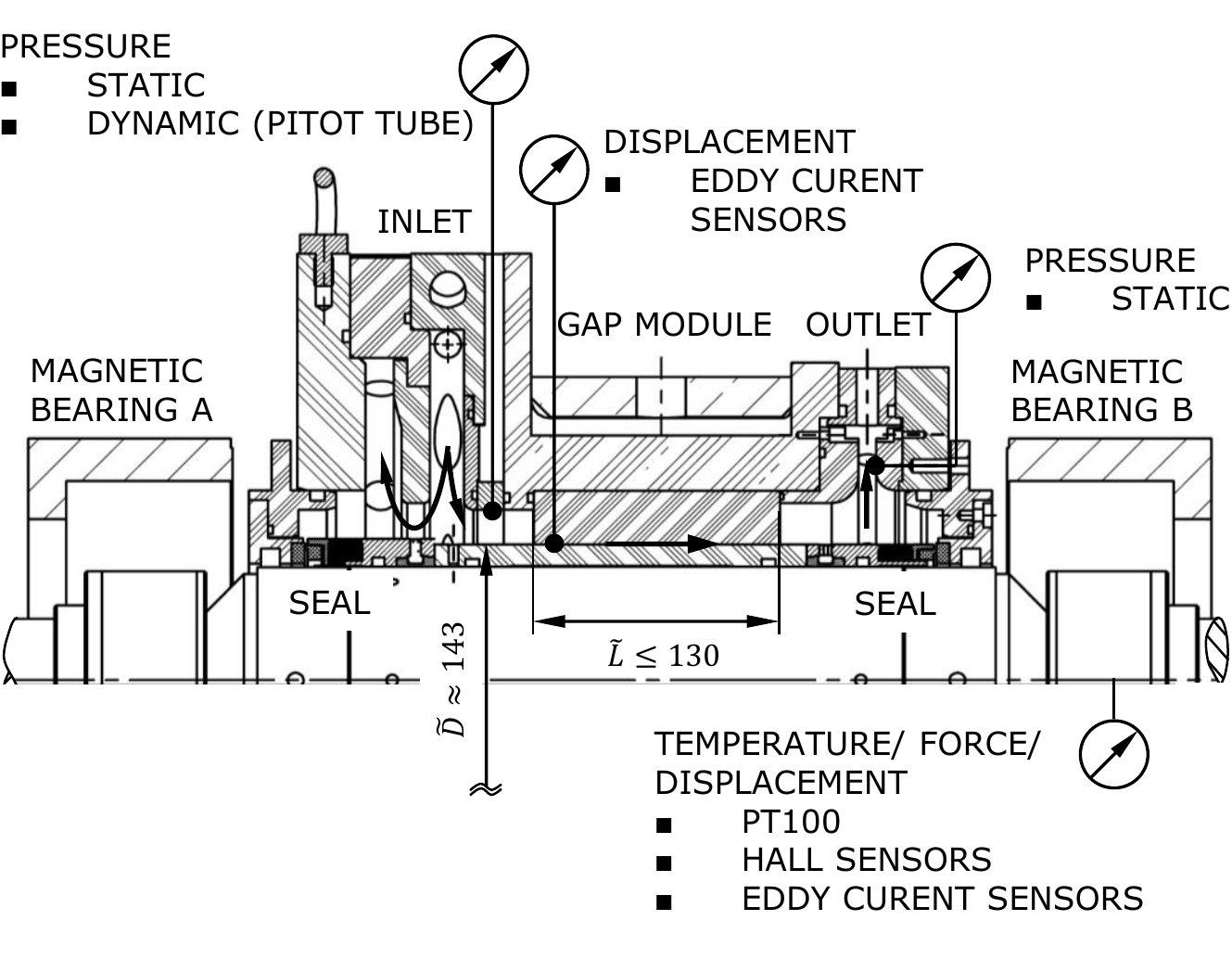}
	\caption{Technical drawing of the gap module.The fluid path is indicated by the black arrows inside the test rig.}
	\label{fig:exp_sketch_test_rig}
\end{figure*}

Figure \ref{fig:exp_sketch_test_rig} shows a technical drawing of the main part of the test rig. The fluid path is indicated by the black arrows. The test rig consists of 5 components: (i) two active magnetic bearings to bear the rotor and measure the induced hydrodynamic forces on the shaft; (ii) the inlet to guide the flow and to measure the pre-swirl as well as the static pressure at the entrance of the annulus; (iii) the gap module; (iv) the outlet and (v) the mechanical seals.

To measure the induced forces on the rotor, each magnetic bearing is equipped with eight hall sensors, one for each pole of each electromagnet. The hall sensors, of type HE144T by Asensor Technology AB ($\delta_{\tilde{B}} = \pm 0.1\% \tilde{B}$), measure the magnetic flux density $\tilde{B}$ in the air gap between the shaft and the magnetic bearing. The force per pole $\tilde{F}_{H,i}$ is quadratically proportional to the magnetic flux (pole surface area $\tilde{A}$, magnetic field constant $\tilde{\mu}_0$)
\begin{equation}
	\tilde{F}_{H,i} = \frac{\tilde{A}}{2\tilde{\mu}_0}\tilde{B}_i, \quad \left(i = 1..8\right).
\end{equation}

Due to the dependence of the magnetic flux density on the position of the rotor inside the magnetic bearing, the hall sensors have to be calibrated. This is done by using an iterative procedure initially developed by Kr\"{u}ger \citep{Kruger.2009}. The calibration is performed using the known rotor mass and its centre of gravity as a reference force $\tilde{F}_{ref}$. For an unloaded shaft, the force measurement of the magnetic bearing with a rotor positioned eccentrically in the bearing must output both the mass as well as the centre of gravity of the rotor. The measurement uncertainty after calibration reduces to $\tilde{\delta}_{\tilde{F},\mathrm{hall}} < \pm 0.035 \tilde{F}$. The displacement of the rotor within the magnetic bearing is measured using four circumferentially distributed eddy current sensors per bearing with an uncertainty of $\tilde{\delta}_{x,\mathrm{AMB}} < \pm 0.75\,\mathrm{\mu m}$. To monitor the temperature of each bearing, each is equipped with two PT100 temperature probes. 

\begin{figure*}
	\centering
	\includegraphics[scale=1.0]{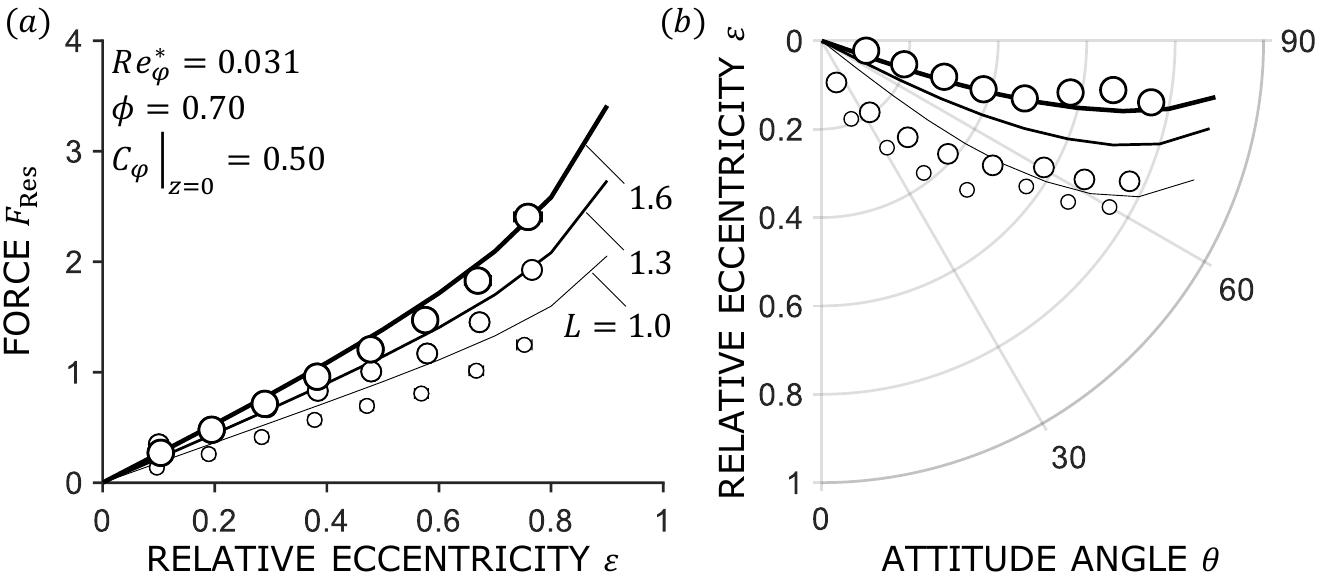}
	\caption{Influence of the annulus length on the (a) resulting force acting on the rotor and (b) the attitude angle.}
	\label{fig:result_force_annulus_length}
\end{figure*}

\begin{figure}
	\centering
	\includegraphics[scale=1.0]{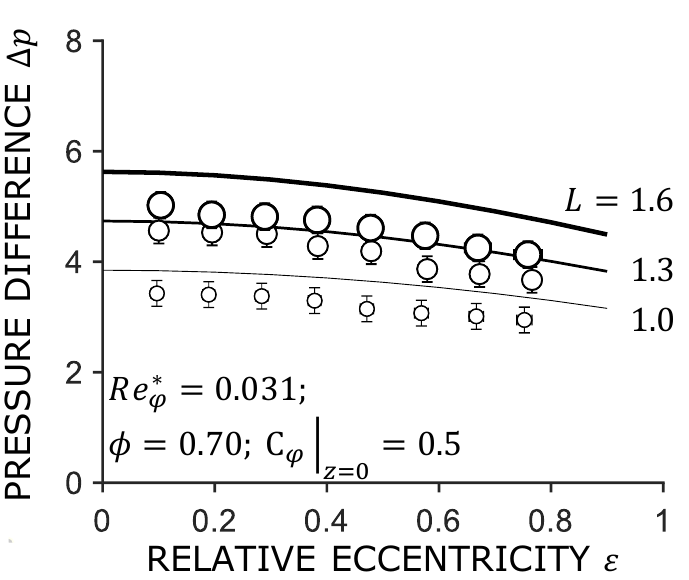}
	\caption{Influence of the annulus length on the pressure difference across the annulus.}
	\label{fig:result_pressure_annulus_length}
\end{figure}

The inlet is specifically designed to generate pre-swirled flows in front of the annulus. It is well known that the pressure field inside the annulus is influenced by the pre-swirl $C_\varphi \big|_{z=0}$. To investigate and quantify the influence on the static characteristics, 12 circumferentially distributed tubes introduce water tangentially into the inlet. By dividing the flow into two parts, the gap and the bypass volume flow, it is possible to continuously vary the circumferential velocity component upstream of the annulus. A Pitot tube connected to a KELLER PD-23/5bar pressure transmitter ($\tilde{\delta}_{\Delta p} < 0.01\, \mathrm{bar}$) is used to measure the circumferential velocity component. The test rig is capable of generating a  pre-swirl in the range of $C_\varphi \big|_{z=0} = 0 \, .. \, 1.7$. In order to determine the position of the rotor within the gap module, the position of the shaft is measured at two planes: (i) at the entrance and (ii) exit of the lubrication gap. For this purpose, two eddylab CM05 eddy current sensors with a user defined measuring range of $1 \, \mathrm{mm}$ and an absolute uncertainty of $\tilde{\delta}_{x,\mathrm{GAP}} < \pm2.4 \, \mathrm{\mu m}$ are used in a $90^\circ$ arrangement. Due to the modular design of the test rig it is possible to vary the length of the annulus as well as the relative clearance by changing the stator inlay and the rotor diameter within the gap module. The relative length can be varied in a range of $0.2 \le L \le 1.8$ and the relative clearance can be modified in a range of $10^{-3} \le \psi \le 10^{-2}$. The supply pressure is measured at the inlet of the gap module by using a KELLER PAA-33X/30bar connected to four wall pressure taps equally space around the annulus with an uncertainty of $\tilde{\delta}_{p,z=0} < \pm 0.033 \, \mathrm{bar}$. The pressure difference across the annulus is measured by a differential pressure sensor KELLER PD-23/20bar with an absolute uncertainty of $\tilde{\delta}_{\Delta p} < \pm0.04 \, \mathrm{bar}$. The test rig is designed to investigate pressure differences of up to $13 \, \mathrm{bar}$. The volume flow through the gap is measured using an ABB DM4311 electromagnetic flowmeter with an absolute uncertainty of $\tilde{\delta}_{\dot{Q},\mathrm{GAP}} < \pm0.04 \, \tilde{Q}_\mathrm{GAP}$. To avoid cavitation within the annulus the test rig can be pressurised up to $15 \, \mathrm{bar}$. The test rig is operated using water at a constant temperature of $35 \, ^\circ\mathrm{C}$ and is fed by a 10 staged $55 \, \mathrm{kW}$ centrifugal pump resulting in flow numbers up to $\phi \le 5$.

\section{Experimental and simulation results}
\begin{figure*}
	\centering
	\includegraphics[scale=1.0]{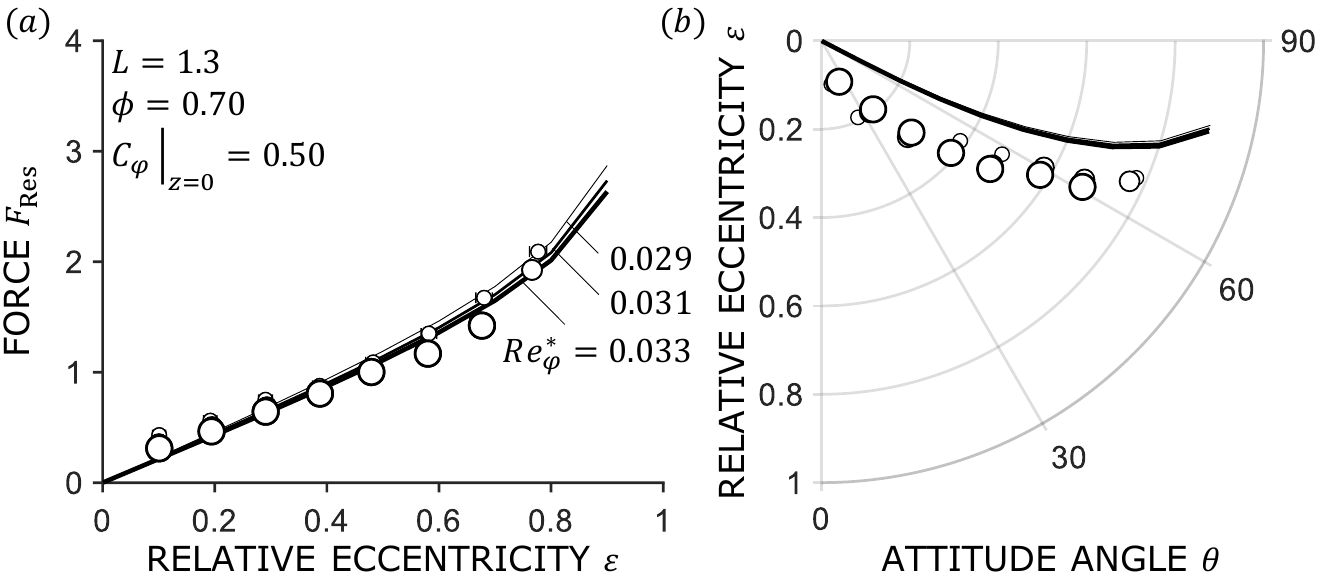}
	\caption{Influence of the modified Reynolds number on the (a) resulting force acting on the rotor and (b) the attitude angle.}
	\label{fig:result_force_modified_reynoldsnumber}
\end{figure*}

\begin{figure}
	\centering
	\includegraphics[scale=1.0]{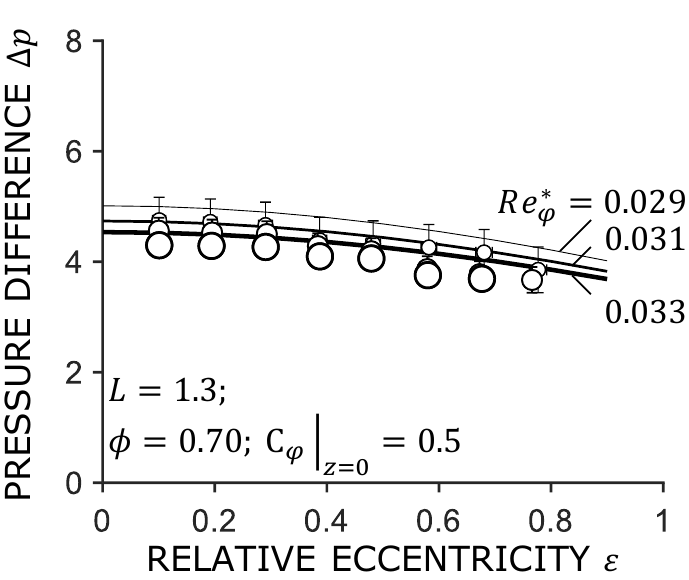}
	\caption{Influence of the modified Reynolds number on the pressure difference across the annulus.}
	\label{fig:result_pressure_modified_reynoldsnumber}
\end{figure}

The experimental investigations are performed using three different annuli with lengths $L =1.0,\,1.3,\,1.6$. Besides the influence of the annulus length, the influence of the modified Reynolds number $Re_\varphi^*$, the influence of the flow number $\phi$ and the influence of the pre-swirl $C_\varphi |_{z=0}$ is investigated. Here, the variation of the modified Reynolds number is achieved by keeping the relative gap clearance constant at $\psi = 4.2 \, \permil$ and modifying the Reynolds number in a range of $Re_\varphi = 3000 ... 5000$. This is reasonable because the Reynolds number and the relative gap clearance only occur as a product in the system of non-linear partial differential equations, cf. chapter \ref{sec:CAPM}. All experiments are conducted over an eccentricity range $0.1 \le \varepsilon \le 0.8$. The experimental results are compared to the ones obtained by the Clearance-Averaged Pressure Model with regard to the resulting force on the rotor $F_{Res}$, the attitude angle $\theta$ and the pressure difference across the annulus $\Delta p$.

For the calculations a fully turbulent flow within the annulus is assumed. The empirical constants of the Hirs' wall friction model are $m_f = 0.0645$ and $n_f = 0.25$, whereas the inlet pressure loss coefficient is $\zeta = 0.25$.

\subsection{Annulus length}
Figure \ref{fig:result_force_annulus_length} (a) shows the influence of the annulus length on the resulting force versus eccentricity. In contrast to the pre validation results, in the following, the lines represent the calculation results obtained by the Clearance-Averaged Pressure Model, whereas the markers represent the experimental data. Again, the line thickness as well as the marker size correlates with the varied parameter. The thicker the line thickness and the larger the marker, the larger the varied  parameter. The modified Reynolds number as well as the flow number and the pre-swirl where kept constant over each measurement within a range of $\pm 1 \, \%$. The influence of the annulus length is investigated choosing a modified Reynolds number $Re_\varphi^* = 0.031$, a flow number $\phi = 0.7$ and a pre-swirl $C_\varphi|_{z=0}$. It is found that the resulting forces on the rotor are in very good agreement with the predicted ones obtained by the CAPM. The force displacement curves is linear at small eccentricities $\varepsilon < 0.5$ and becomes increasingly non-linear with increasing eccentricity. This is due to the fact that the induced forces at small eccentricities are mainly caused by the Lomakin effect, while the forces induced by the hydrodynamic effect become more dominant as the eccentricity increases.

Figure \ref{fig:result_force_annulus_length} (b) shows the corresponding attitude angle versus eccentricity. It shows good agreement between the measured and predicted attitude angle, with the differences between measured and calculated results decreasing with increasing length. Here, the attitude angle increases with increasing annulus length. In contrast to a classical journal bearing the attitude angle does not follow the Gümbel curve, i.e. a semicircular displacement of the rotor loci with increasing eccentricity. This is reasonable because cavitation is not considered in the CAPM and it is prevented during the tests by increasing the pressure level of the entire test rig. The prevention of cavitation leads to a resulting force dominated by the $Y$ force component, resulting in a mainly horizontally moved rotor.

Figure \ref{fig:result_pressure_annulus_length} shows the pressure difference across the annulus versus eccentricity. It shows a decreasing pressure difference with increasing eccentricity and an increasing pressure difference with increasing annulus length. The decreasing pressure difference with increasing eccentricity is due to the fact that the flow resistance opposed by the annulus decreases with increasing eccentricity.

\subsection{Modified Reynolds number}
\begin{figure*}
	\centering
	\includegraphics[scale=1.0]{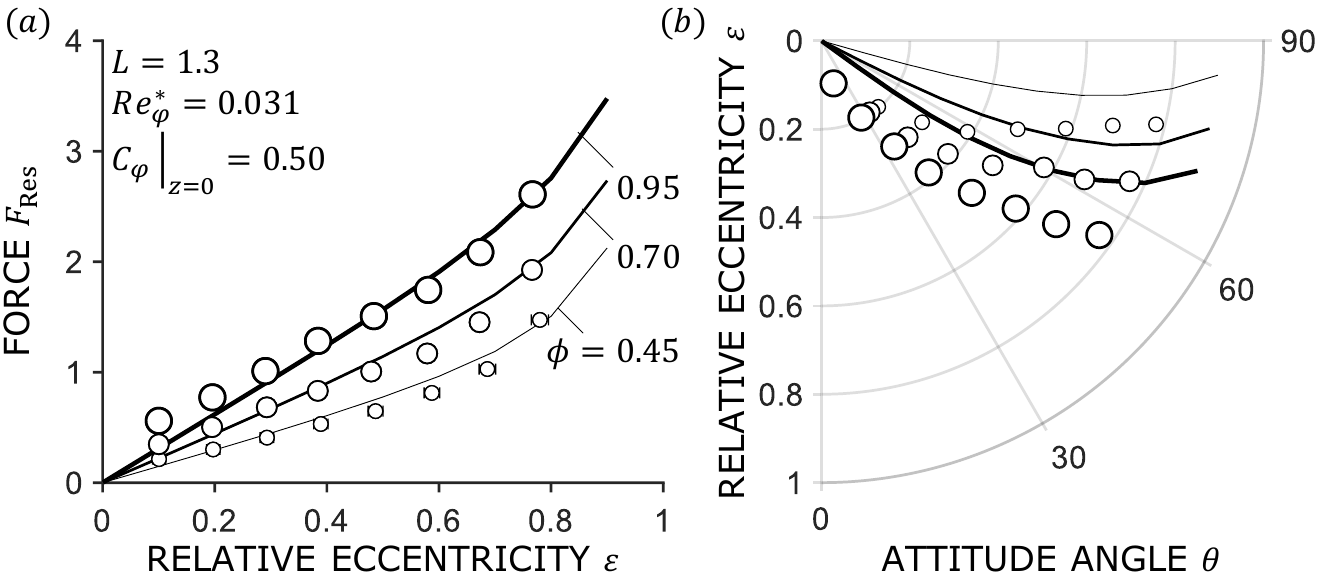}
	\caption{Influence of the flow number on the (a) resulting force acting on the rotor and (b) the attitude angle.}
	\label{fig:result_force_flow_number}
\end{figure*}

\begin{figure}
	\centering
	\includegraphics[scale=1.0]{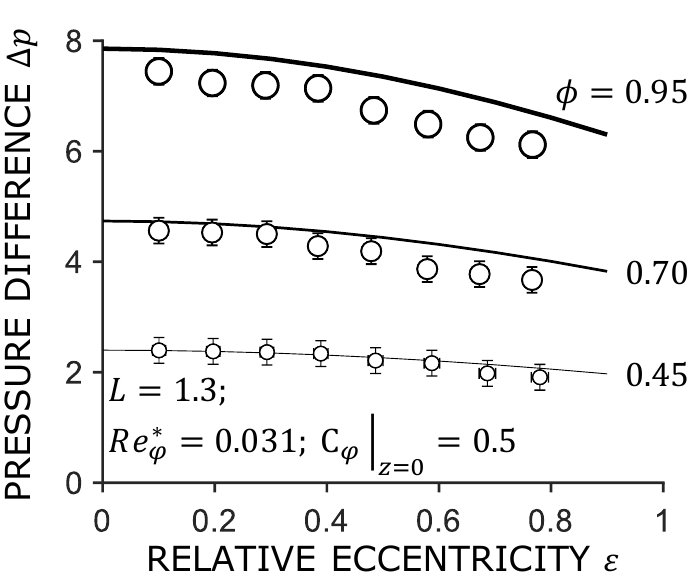}
	\caption{Influence of the flow number on the pressure difference across the annulus.}
	\label{fig:result_pressure_flow_number}
\end{figure}
In the following the influence of the modified Reynolds number on the force, attitude angle and pressure difference is investigated. Figure \ref{fig:result_force_modified_reynoldsnumber} (a) shows the influence of the modified Reynolds number on the resulting force on the rotor versus eccentricity for an annulus length $L=1.3$, a flow number $\phi=0.7$ and a pre-swirl $C_\varphi|_{z=0}=0.5$. Due to the constant clearance $\psi= 4.2\,\permil$ the modified Reynolds number is controlled by controlling the angular frequency of the rotor. Here, the three modified Reynolds numbers $Re_\varphi^*= 0.029$, $0.031$ and $0.033$ correspond to the three Reynolds numbers~$Re_\varphi = 3000$, $4000$ and $5000$. Figure \ref{fig:result_force_modified_reynoldsnumber} (a) shows a good agreement between the experimental results and the predictions of the Clearance-Averaged Pressure Model. By increasing the modified Reynolds number the force on the rotor slightly decreases. At first this seams contradictory but by recapping the dimensionless force $F_\mathrm{Res} := 2 \tilde{F}_\mathrm{Res} / \left( \tilde{\varrho} \tilde{\Omega}^2 \tilde{R}^2 \tilde{R} \tilde{L} \right)$ becomes quite clear because of the fact that the force is inversely proportional to $F \propto 1 / \tilde{\Omega}^2$. 

Figure \ref{fig:result_force_modified_reynoldsnumber} (b) shows the influence of the modified Reynolds number on the attitude angle versus eccentricity. Here, the attitude angle predicted by the CAPM slightly decreases by increasing the modified Reynolds number. However, due to the small changed in angle the experimental data does not show a clear trend towards a decreasing or increasing attitude angle. Nevertheless, the experimental data and the simulations show a good agreement. 

Figure \ref{fig:result_pressure_modified_reynoldsnumber} shows the influence of the modified Reynolds number on the pressure difference across the annulus versus eccentricity. It shows a decreasing curve with increasing eccentricity. In addition, by increasing the modified Reynolds number the pressure difference decreases. This is due to the fact that the dimensionless pressure difference $\Delta p := 2 \Delta \tilde{p} / \left( \tilde{\varrho} \tilde{\Omega}^2 \tilde{R}^2 \right)$ is also inversely proportional to the square of the angular frequency of the rotor $\Delta p \propto 1 / \tilde{\Omega}^2$.

\subsection{Flow number}
In the following the influence of the flow number on the force, the attitude and the pressure difference across the annulus is investigate. Figure \ref{fig:result_force_flow_number} (a) shows the force versus eccentricity for an annulus length $L= 1.3$, a modified Reynolds number $Re_\varphi^*=0.031$ and a pre-swirl of $C_\varphi|_{z=0}=0.5$. It shows a good agreement between the simulation results an the experimental data. The induced force on the rotor is increasing by increasing the flow number. This is reasonable because an increase in flow number results in an increased pressure difference across the annulus, cf. figure \ref{fig:result_pressure_flow_number}. Therefore, the Lomakin effect is increased resulting in an increased resulting force on the rotor. 

Figure \ref{fig:result_force_flow_number} (b) shows the corresponding attitude angle versus eccentricity. The predictions of the CAPM are again in good agreement with the experimental data. Here, the attitude angle decrease when the flow number is increased. This is reasonable, as mentioned before, the prevention of cavitation inside the annulus leads to a dominating $Y$ force component acting on the rotor. Therefore, the rotor is mainly moved horizontally. The increased Lomakin effect due to the increasing flow number results in an increasing force on the rotor direction pointing to the annulus centre, i.e. a force in negative $X$ direction. Therefore, the attitude angel decreases with an increasing flow number.

Figure \ref{fig:result_pressure_flow_number} shows the influence of the flow number on the pressure difference across the annulus. As mentioned above, by increasing the flow number the pressure difference increases, resulting in an increased Lomakin effect. The experimental data are in good agreement with the results calculated by the Clearance-Averaged Pressure Model.

\subsection{Pre-swirl}
\begin{figure*}
	\centering
	\includegraphics[scale=1.0]{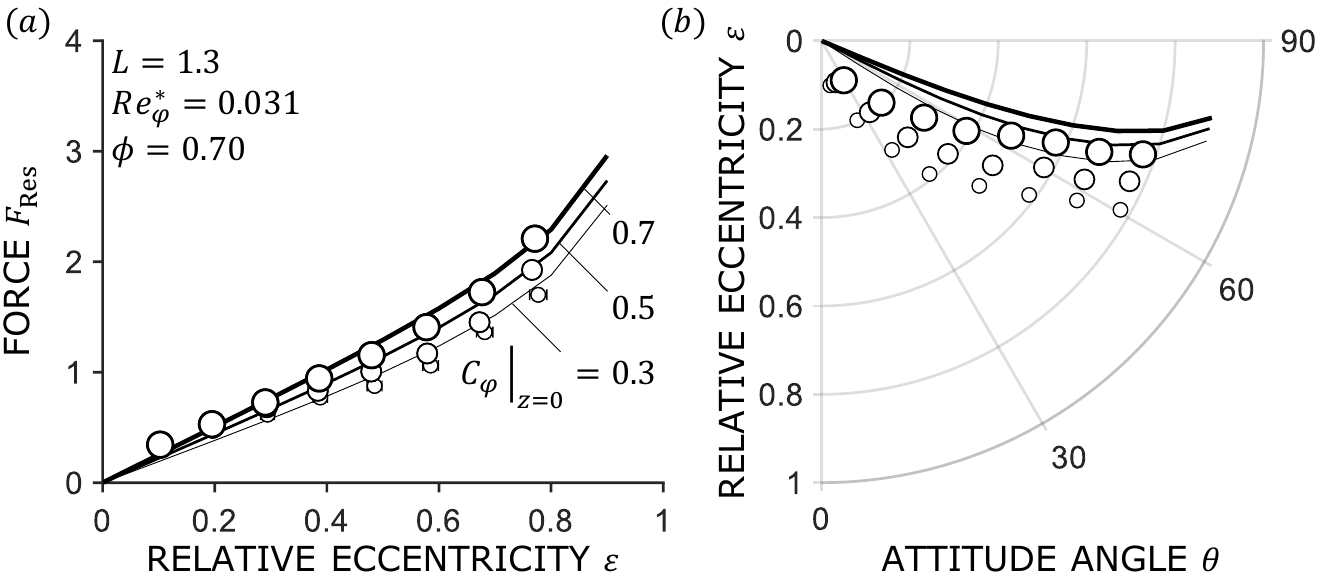}
	\caption{Influence of the pre-swirl on the (a) resulting force acting on the rotor and (b) the attitude angle.}
	\label{fig:result_force_pre_swirl}
\end{figure*}

Finally, the influence of the pre-swirl on the force on the rotor, the attitude angle and the pressure difference across the annulus is investigated. Figure \ref{fig:result_force_pre_swirl} (a) shows the influence of the pre-swirl on the resulting force for an annulus length $L=1.3$, a modified Reynolds number $Re_\varphi^*=0.031$ and a flow number $\phi = 0.7$. Both, the experimental data and the simulations by the CAPM show an increasing force due to the increased pre-swirl in front of the annulus. This is reasonable because an increase in pre-swirl mainly increases the force in circumferential direction, i.e. $F_Y$, resulting in an increased resulting force and an increasing attitude angle. Figure \ref{fig:result_force_pre_swirl} (b) shows the attitude angle versus eccentricity. As mentioned, the angle is increased by increasing the pre-swirl in front of the annulus. The predicted attitude angles are in good agreement with the experimental results.

\begin{figure}
	\centering
	\includegraphics[scale=1.0]{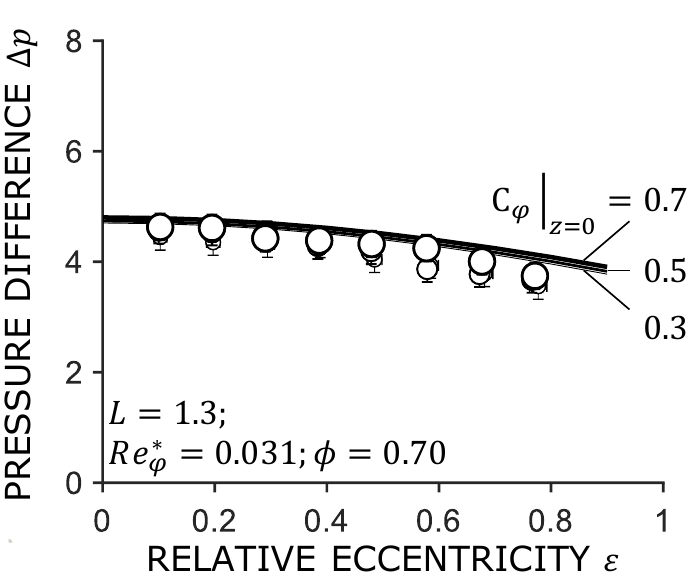}
	\caption{Influence of the pre-swirl on the pressure difference across the annulus.}
	\label{fig:result_pressure_pre_swirl}
\end{figure}

Figure \ref{fig:result_pressure_pre_swirl} shows the influence of pre-swirl on the pressure difference across the annulus. Here, the CAPM predicts a slight increase in the pressure difference, whereas a distinct tendency in the measurement data is not apparent. The deviations between the measurements are within the measurement uncertainty. The increase predicted by the model is reasonable, since the increase in circumferential velocity leads to an increased friction in the annulus, which in return results in an increased pressure difference across the annulus.  

In summary, the following facts and reasons can be derived from the simulations and experiments:
\begin{itemize}
    \item[(i)] The experimental data are in good agreement with the simulations performed by the Clearance-Averaged Pressure Model.
    \item[(ii)] By increasing the annuls length the force on the rotor, the attitude angle and the pressure difference across the annulus is increased. 
    \item[(iii)] By increasing the modified Reynolds number the force on the rotor, the attitude angle and the pressure difference is decreased.
    \item[(iv)] By increasing the flow number the Lomakin effect increases, resulting in an increased force on the rotor and and increased pressure difference across the annulus. In contrast, the attitude angle decreases when the flow number is increased.
    \item[(v)] By increasing the pre-swirl the force on the rotor, the attitude angle and the pressure difference are increased.  
\end{itemize}

\section{Conclusions}
In the presented, the static force characteristics of annular gaps with an axial flow component are discussed. First, the state-of-the-art modelling approaches to describe the flow inside the annulus are recapped and discussed. The discussion focuses in particular on the modelling of inertia effects. Second, a new calculation method, the Clearance-Averaged Pressure Model (CAPM) is presented. The CAPM used an integro-differential approach in combination with power law ansatz functions for the velocity profiles and a Hirs’ model to calculate the resulting pressure field. Due to the modular integration of the velocity profiles, the ansatz functions can be adapted to gap flows at arbitrary Reynolds numbers. The requirement $Re_\varphi \to \infty$ for block-shaped velocity profiles in the bulk-flow approach is thus eliminated. In particular, the power law ansatz functions can be used to treat both laminar and turbulent gap flows in a unified model framework. Third, an experimental setup is presented using magnetic bearings to inherently measure the position as well as the force induced by the flow field inside the gap. Compared to existing test rigs where the shaft is supported by ball or journal bearings, magnetic bearings have the advantage of being completely contactless and thus frictionless. In addition, the ability to displace and excite the shaft at user-defined frequencies makes them ideal for determining the static and dynamic characteristics of annular gap flows. Fourth, an extensive parameter study is carried out focusing on the characteristic load behaviour, attitude angle and pressure difference across the annulus. The experimental results are compared to the results of the Clearance-Averaged Pressure Model. It is shown that by increasing the annuls Length the force on the rotor, the attitude angle and the pressure difference across the annulus increase. In addition, by increasing the modified Reynolds number the force on the rotor, the attitude angle and the pressure difference decrease. Due to the Lomakin effect, increasing the flow number will result in an increased force on the rotor as well as an increased pressure difference. In contrast, the attitude angle decreases. By increasing the pre-swirl upstream of the annulus the force on the rotor, the attitude angle and the pressure difference increase.  

\section*{Acknowledgements}
We gratefully acknowledge the financial support of the industrial collective research programme (IGF no. 19225/BG 1), supported by the Federal Ministry for Economic Affairs and Energy (BMWi) through the AiF (German Federation of Industrial Research Associations e.V.) based on a decision taken by the German Bundestag. In addition, we kindly acknowledge the financial support of the Federal Ministry for Economic Affairs and Energy (BMWi) due to an enactment of the German Bundestag under Grant No. 03ET7052B and KSB SE \& Co. KGaA. Special gratitude is expressed to the participating companies and their representatives in the accompanying industrial committee for their advisory and technical support. 

\section*{Declaration of competing interest}
The authors declare that they have no known competing financial interests or personal relationships that could have appeared to influence the work reported in this paper.

\bibliographystyle{asmems4}
\bibliography{main-refs}

%----------------------------------------------------------------------------------------

\end{document}